\documentclass[useAMS,usenatbib]{mn2e}
\pdfoutput=1 
\usepackage[pdftex]{graphicx} 
\usepackage{amssymb}
\newcommand{\noteA}[1]{#1} 
\newcommand{\note}[1]{#1} 

\title[Applying a one-dimensional PDR model to Taurus]{Applying a one-dimensional PDR model to the Taurus molecular cloud and its atomic envelope}
\author[J. S. Heiner \& E. V\'{a}zquez-Semadeni]{J. S. Heiner$^{1}$\thanks{E-mail: j.heiner@crya.unam.mx}, E. V\'{a}zquez-Semadeni$^{1}$\\
  $^1$Centro de Radioastronom\'{i}a y Astrof\'{i}sica (CRyA), Universidad Nacional Aut\'{o}noma de M\'{e}xico, C.P. 58190 Morelia, Michoac\'{a}n, Mexico}

\begin{document}

\date{}
\pagerange{\pageref{firstpage}--\pageref{lastpage}} \pubyear{2012}
\maketitle
\label{firstpage}

\begin{abstract}
  In this contribution, we \noteA{test our previously published} one-dimensional PDR model \noteA{for deriving} \textit{total} hydrogen volume densities from HI column density measurements \noteA{in extragalactic regions by applying it to the Taurus molecular cloud, where its predictions can be compared to available data. Also,} we make the first direct detailed comparison of our model to CO(1-0) and far-infrared emission.

  Using an incident UV flux $G_0$ of 4.25 ($\chi = 5$) throughout the main body of the cloud, we derive total hydrogen volume densities of $\approx 430\ \rm{cm}^{-3}$, consistent with the extensive literature available on Taurus. The distribution of the volume densities shows a log-normal shape with a hint of a power-law shape on the high density end. We convert our volume densities to $\rm{H_2}$ column densities assuming a cloud depth of 5 parsec and compare these column densities to observed CO emission. We find a slope equivalent to a CO conversion factor relation that is on the low end of reported values for this \note{factor} in the literature ($0.9 \times 10^{20}\ \rm{cm^{-2} (K~km~s^{-1})^{-1}}$), although this value is directly proportional to our assumed value of $G_0$ as well as the cloud depth. We seem to under-predict the total hydrogen gas as compared to 100 $\mu$m dust emission, which we speculate may be caused by a higher actual $G_0$ incident on the Taurus cloud than is generally assumed.
\end{abstract}


\section{Introduction}

Molecular hydrogen is widely considered the key ingredient in star formation. However, on extragalactic scales it is only observed directly at higher densities or excitation conditions. Photodissociation regions (where molecular hydrogen is destroyed) are key to learning more about the molecular gas, because observable amounts of atomic hydrogen are produced in these regions. 

\note{The available literature on photodissociation regions (PDRs) is extensive. Early work includes \citet{1971ApJ...163..165H} and \citet{1974ApJ...191..375J} on the formation and destruction of molecular hydrogen, \citet{1975ApJ...197..347G} on the appearance of $\rm{C^{+}}$ and CO, and calculations on more atomic and molecular species by e.g. \citet{1977ApJS...34..405B} and \citet{1986ApJS...62..109V}. Recent years have seen the arrival of various publicly available computer models that aim to solve the detailed structure of PDRs for many atomic and molecular species, such as CLOUDY \citep{1998PASP..110..761F}, the Meudon PDR code \citep{2006ApJS..164..506L}, the KOSMA-$\tau$ code \citep{2006A&A...451..917R} and the 3D-PDR code \citep{MNR:MNR22077}. A variety of such codes were compared and the results published by \citet{2007A&A...467..187R}. For the focus of this work, we will limit our attention to atomic and molecular hydrogen (as well as, to a certain extent, carbon monoxide) in low density molecular clouds, since we aim to explore and provide an independent probe of total hydrogen volume densities as opposed to a detailed model of the internal structure that includes predicted line intensities of various species.}

In lower-density, lower-radiation regimes, it is possible to derive total hydrogen densities using the relatively straightforward process of photodissociation of molecular hydrogen, which results in detectable atomic hydrogen emission\note{, potentially on galactic scales}. A detailed study of a one-dimensional approach to such photodissociation regions was carried out by \citet{1991ApJ...377..192H}, including calculations connecting gas densities with the incident UV flux, CO and 100 $\mu$m dust emission. Their results have been used to model the interstellar medium of external galaxies as well as the Milky Way.


\citet{2002ApJ...568..242L} studied PDRs in what they called the intermediate regime, or $n \approx 500-5000\ \rm{cm}^{-3}$ and $G_0 \approx 10-60$, in the L1204 and S140 regions. While they successfully predicted several atomic and molecular emission lines, they observed enhanced $\rm{H_2}$ emission that could possibly be explained by enhanced formation rates or weak shocks related to turbulent decay.

The notion that a large fraction of extragalactic atomic hydrogen in galactic disks is produced in PDRs has been around for quite some time \citep{1986Natur.319..296A,1997ApJ...487..171A,2000ApJ...538..608S,2011MNRAS.416....2H}.  Using PDR-produced atomic hydrogen, it becomes possible to calculate total hydrogen volume densities at selected positions across galactic disks using a simple \note{approximate analytical} one-dimensional PDR model \citep[see also e.g.][]{2011ApJ...739...97H}. The resulting densities are consistent with values expected for the diffuse interstellar medium. Corresponding molecular emission (CO) is found near higher (several hundred cm$^{-3}$) total hydrogen volume densities, although it should be noted that under-excited CO could still hide \note{in} relatively dense clouds. The densities obtained in this manner are also consistent with the (volumetric) Schmidt Law, linking the local star formation rate to the total hydrogen volume density \citep[][]{2010ApJ...719.1244H}, at least in the power-law exponent. 

However, so far this approach has not been compared directly to CO-derived molecular hydrogen gas contents in a detailed fashion. In order to do this, we turn to our own Galaxy to apply the one-dimensional PDR model to a molecular cloud that has densities in the range where this model can be expected to apply, namely below $10^4\ \rm{cm^{-3}}$.

We selected the well-known and heavily-studied Taurus molecular cloud, part of the Perseus-Taurus-Auriga complex for this study. It has molecular gas densities of a few hundred cm$^{-3}$ with denser clumps \citep[for a recent comprehensive study, see e.g.][]{2010ApJ...721..686P} and is relatively quiescent in terms of star formation. 

The atomic hydrogen in the Taurus area is considered to be connected to the molecular gas \citep{2003ApJ...585..823L}. Several surveys of Taurus have been completed in the CO \mbox{1-0} line \citep{1987ApJS...63..645U,2008ApJS..177..341N} as well as a number of studies focusing on specific regions.

According to \citet{1987ApJS...63..645U}, the Taurus CO mass is of the order of $10^4\ \rm{M_\odot}$, depending on which parts of the complex are included \citep[see also][]{2010ApJ...721..686P}. Furthermore, clouds in the Taurus complex \noteA{were found to have CO-derived masses approximately equal to their virial masses} \citep{1987ApJ...321..855H}. \citet{1995A&AS..111..483A} compared the Taurus CO emission with IRAS far-infrared emission maps to conclude that the CO emission should trace at least 70\% of the matter indicated by the 100$\mu$m emission, using the \citet{1987ApJS...63..645U} maps. This makes Taurus a good candidate to compare the results of the simple one-dimensional PDR model to known molecular data, which will support the case for applying this model in the extragalactic case as well.

In the case of Taurus, then, we have a molecular cloud that is bathed in a photodissociating flux coming from the outside. \note{We may assume that the molecular cloud is surrounded by an envelope of atomic hydrogen, that is either in the process of becoming molecular, or originates from photodissociated molecular hydrogen. Either way, this atomic hydrogen is part of the balance of photodissociation.}
Our study provides a starting point to further investigation of the atomic-to-molecular transition around molecular clouds in the context of the kinematics of the gas, as prompted by, e.g., the work presented by \citet{1999ApJ...527..285B} and \citet{2011A&A...529A..41N}.

The application of the simple one-dimensional PDR model includes an assumption about the details of the conversion of HI to $\rm{H_2}$ and back, namely that (at least in the extragalactic case) the full integrated HI column can be assumed to be part of the process, minus a local background value. For an edge-on large-scale PDR, this assumption is likely to overestimate the HI column and therefore underestimate the related total hydrogen volume density. \note{The reason for this is that the line of sight from the observer through the galaxy probes edge-on HI shells more or less parallel to the shell's surface, which would yield a higher column density than the column density seen at a right angle to the surface (assuming a shell that is thinner than it is wide). The true three-dimensional structure of the cloud remains unknown, which is especially true for the extragalactic case.}

In our own Galaxy, \citet{2004ApJ...608..314A} used a limited velocity range dictated by the local molecular emission (for Maddalena's Cloud), an approach that we also adopt here. Under these assumptions, we can trivially calculate the fraction of the total HI column density that is included in the chosen velocity range. However, this in itself only hints at the fraction of HI that is part of the transition from atomic to molecular gas (and back).

Also, while \citet{2004ApJ...608..314A} applied the simple PDR model to Maddalena's Cloud in our own Galaxy, a detailed, \note{spatially} resolved application fell outside the scope of their work. With the application of this method to Taurus, then, we expand on their result. Consequently, this allows us to study the distribution of total hydrogen volume densities in the Taurus region. This distribution holds important clues to the dynamical state of the cloud. 

This paper is structured as follows:
First, we briefly review the theory behind the one-dimensional PDR model and how it can be applied to the case of Taurus. We then describe the archival data we used to explore the Taurus region. We discuss obtaining appropriate atomic hydrogen column densities, after which we present total hydrogen volume density maps of the main Taurus area. We then proceed to compare our results to CO emission data, as well as far-infrared data. We check our results for consistency with the \citet{2004ApJ...608..314A} calculations and discuss the values of the UV field suggested by our study. Finally we end with a summary of our conclusions.

\section{A simple one-dimensional PDR model applied to the molecular cloud-atomic envelope interface}
\label{sec:theory}


We use a straightforward one-dimensional plane-parallel PDR model that connects atomic hydrogen, molecular hydrogen (through the total hydrogen volume density), the incident UV flux and the local dust-to-gas ratio. For a more thorough description we refer to, e.g., \citet{2008ApJ...673..798H}. The final\note{, approximate} formula \note{describing} the photodissociation balance, which we employ here, is:
\begin{equation}
  n = 106\ G_0 \left(\frac{\delta}{\delta_0}\right)^{-0.3}\left[\exp{\left(\frac{N_{\rm{HI}}(\delta/\delta_0)}{7.8\times 10^{20}}\right)}-1\right]^{-1},
  \label{eqn:ntot}
\end{equation} 
where $n$ is the total hydrogen volume density in $\rm{cm}^{-3}$ \note{averaged along the line of sight through the cloud}, $G_0$ is the incident UV flux and $\delta/\delta_0$ is the dust-to-gas ratio scaled to the solar neighborhood value and $N_{\rm{HI}}$ is the (PDR-produced) HI column density in the direction of the incident UV flux. The exponent in $\delta/\delta_0$ is slightly different from the one that was used previously because it includes the improvements suggested by \citet{2009ApJ...694..978H}. 

\note{The atomic hydrogen column density predicted by this approximate analytical model is scaled to fit the column densities produced by numerical PDR calculations presented by \citet{1999ApJ...527..795K}, as described in \citet{2004ApJ...608..314A}. It assumes that the formation and destruction of molecular hydrogen are balanced, at least when integrated through the gas cloud \citep{1988ApJ...332..400S}. It provides an estimate of the total hydrogen volume density at the location where a certain HI column density is measured, neglecting the finer details of the formation and destruction of molecular hydrogen and other elements in order to focus on the direct, line-of-sight-averaged connection between atomic hydrogen and molecular hydrogen.}

In the current context, $G_0 = 0.85 \chi$ \citep[][]{1999RvMP...71..173H}, where $G_0$ is the incident radiation in units of the Habing flux \citep{1968BAN....19..421H} and $\chi$ is in units of the Draine field \citep{1978ApJS...36..595D}. \note{These definitions of the radiation field and their relation are valid under the assumption of a cloud approximated by a semi-infinite slab, illuminated isotropically, with radiation therefore entering through $2 \pi$ steradians. In the case of Taurus, it is not clear where the incident flux originates, and we approximate the ambient radiation field by adopting a constant value for it following \citet{2010ApJ...715.1370G}. Equation (\ref{eqn:ntot}) requires a single value for the incident flux, namely the one at the edge of the cloud. In our model, it is not necessary to know whether the radiation originates from in front of, or from behind the cloud. The value of $n$, however, is an average and includes all the hydrogen gas along the line of sight through the cloud. This includes the atomic hydrogen at the edge of the cloud as well as the molecular hydrogen further inwards. This average value may include clumped gas along the line of sight as well.}

\noteA{This model is thought to be most useful in the density range $10\ \rm{cm}^{-3} \lesssim n \lesssim 10^5\ \rm{cm}^{-3}$ and $G_0 \lesssim 100$ \citep{2004ApJ...608..314A}. The underlying assumption is that (a large fraction of the) measured HI column density is indeed produced by photodissociation due to the nearby source of UV flux.}
Trivially, the total hydrogen column density $N = n \times d_{\rm{cloud}}$ where $d_{\rm{cloud}}$ is the assumed cloud depth. Throughout this paper, $n$ or $N$ without subscript will refer to total hydrogen quantities, that is $N = N_{\rm{HI}} + 2 N_{\rm{H_2}}$.

In the extragalactic case, one needs to identify patches of HI in column density maps that are suspected to be associated with central OB star associations as seen on far-UV images. The linear dimension of the PDR model is then taken to stretch on the plane of the sky from the central star cluster through the HI patch. From the projected angular separation, a linear distance and a corresponding incident flux on the location of the HI emission are then calculated. \note{It is important to select a PDR that is seen approximately edge-on in order to be able to calculate the incident flux, although we will suffer from projection effects. While the HI column density is not measured in the direction from the star cluster through the HI patch, it is nevertheless assumed that the measured column density is an appropriate probe of the required column density, as we explained in the Introduction. Specifically, we refer the reader to \citet{2008ApJ...673..798H}, their Figure 4c, showing an idealized representation of the expected morphology. Formally, a spherical molecular cloud is assumed, or a tubular filament in the plane of the sky. A more thorough study of the projection effects on the measured HI column density in the extragalactic case falls outside the scope of this paper, and will be investigated in more detail in a future contribution.}

\note{The applicability of the method} is limited by the angular resolution of the images, meaning that only large-scale PDRs can be used (about 50-500 parsec separation). To apply this same method to PDRs in our own Galaxy, we take a slightly different approach. Instead of considering an edge-on PDR, we look at the PDR face-on. Here, we consider a uniform radiation field impinging on a known molecular cloud along the line of sight (with the radiation coming either from the direction of the observer or from behind the cloud), and attempt to isolate the photodissociated atomic hydrogen in front and behind the molecular cloud. The linear dimension of the PDR model is therefore perpendicular to the plane of the sky, and we can consider each pixel on the observed map individually, as long as the approximation of the incident radiation going along the line of sight holds. \note{In principle, we should now measure the HI column density in the proper direction, although the three dimensional structure of the cloud remains unknown.}

\begin{figure*}
  \resizebox{1.5\columnwidth}{!}{\includegraphics{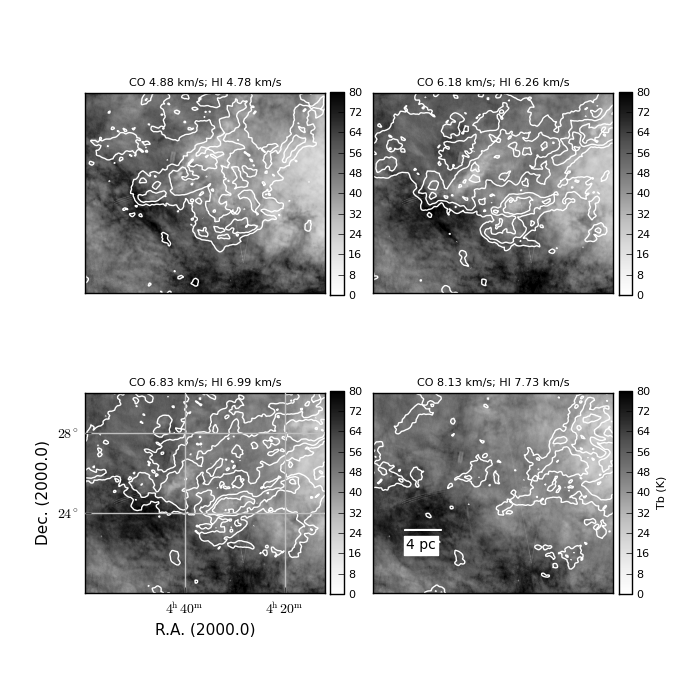}}
  \caption{\label{fig:hico567} HI and CO emission near 5, 6, 7 and 8 km s$^{-1}$. CO contours are main beam brightness temperatures $T_{mb}$ of 1, 3.5 and 6 K.}
\end{figure*}

\begin{figure}
  \resizebox{\columnwidth}{!}{\includegraphics{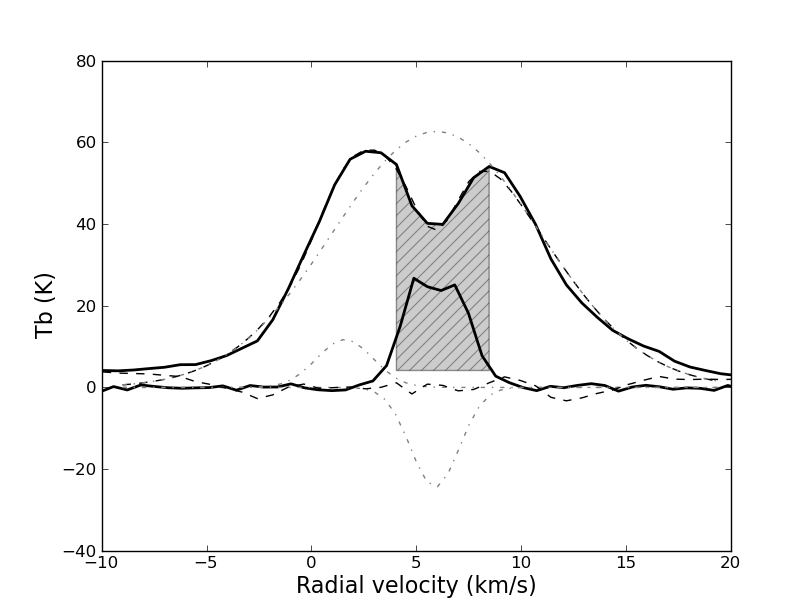}}
  \caption{\label{fig:profile_ex} Sample profile from the GALFA HI datacube of Taurus. This particular profile shows a clear absorption feature as shown by three gaussians used to fit the line. Thick solid lines: HI profile and CO profile (multiplied by a factor of 5 for clarity). Dash-dotted lines: gaussian components of the fit. Dashed lines: fitted profile and residual. Gray area: section of the profile used to infer the local HI column density.}
\end{figure}

In this paper, we will compare our predicted column density $N_{\rm{H_2}}$ directly to the observed (velocity-integrated) $W_{\rm{CO}}$ CO(1-0) line emission. This CO emission is routinely converted to $\rm{H_2}$ column densities using a constant conversion factor, otherwise known as $X_{\rm{CO}}$. In principle this quantity is defined only for an ensemble of molecular clouds on a galactic scale, but this conversion has also been applied on sub-kpc scales \citep{2008AJ....136.2846B} or specific regions of our Galaxy \citep{2010ApJ...710..133A}. The value of this so-called X-factor is thought to range between $0.9 - 4\ \times 10^{20}\ \rm{cm^{-2}\ (K\ km\ s^{-1})^{-1}}$, but a Galactic standard value in the same units is usually adopted for any galaxy in the absence of more direct measurements of this value. \noteA{For example, \citet{1998ApJ...498..541K} used a factor of $2.8\ \times 10^{20}\ \rm{cm^{-2}\ (K\ km\ s^{-1})^{-1}}$ for a sample of 61 galaxies to derive $\rm{H_2}$ column densities.}  The lower values have come from e.g. \citet{1996ApJ...463..609D} and the higher values from e.g. \citet{2007ApJ...663..866D}.

For the purpose of our comparisons we will use the CO X-factor value adopted by \citet{2001ApJ...547..792D}, namely $1.8\ \times 10^{20}\ \rm{cm^{-2}\ (K\ km\ s^{-1})^{-1}}$, since we use their CO data in this work. It should be noted that we use this value for reference purposes only, as it is helpful to plot this value as the slope in our figures comparing $N_{H_2}$ to $W_{CO}$. At our parsec and sub-cloud scale a constant conversion factor should not be expected.

\section{Data and parameters}

For the one-dimensional PDR model described in Equation (\ref{eqn:ntot}) we need the following parameters: the incident UV flux $G_0$, HI column densities, and the dust-to-gas ratio (scaled to the solar neighborhood) $\delta/\delta_0$. In addition we will be comparing our predicted densities to CO(1-0) and far-infrared (100 $\mu \rm{m}$) derived $\rm{H_2}$ column densities.

Following \citet{2010ApJ...715.1370G}, we adopt a uniform incident $\chi$ of 5 (or $G_0 = 4.25$). They consider this value on the high end of the range based on a census of nearby B-stars, but they require at least a value this high to explain their observations. In our model, $n \propto G_0$, which means that underestimating $G_0$ also means underestimating the total hydrogen volume density.

We assume a solar dust-to-gas ratio ($\delta/\delta_0 = 1$). Fixing both the incident UV flux $G_0$ and the dust-to-gas ratio, appropriate for a relatively small region like the Taurus molecular cloud (as opposed to large-scale PDRs up to several hundred parsec across as investigated in the extragalactic case), means that the predicted total hydrogen density will vary with the HI column density only.

We use the 21-cm data from the GALFA-HI survey which has an effective beam size of $3.9 \times 4.1$ arcmin \citep{2011ApJS..194...20P}. We retrieved the data with a velocity separation of 736 m s$^{-1}$, ranging from a radial velocity of -10 km s$^{-1}$ to 20 km s$^{-1}$, which is more than enough to include the Taurus molecular cloud at a radial velocity between 5 and 6 km s$^{-1}$.

\citet{2009MNRAS.395L..81B} modeled the main part of the molecular cloud as a  $32 \times 5$ pc filament at a distance of about 145 pc, based on the distance measurements of \citet{2007ApJ...671..546L} and \citet{2007ApJ...671.1813T}. \noteA{We will adopt this model here.} The actual distance does not influence our results. 

We use the CO data of Taurus presented by \citet{2001ApJ...547..792D} for comparison. We reprojected it to the HI map first, but this results in an over-sampled CO map. We therefore smoothed the HI data to the CO resolution of 8.4 arcmin. Then, we regridded the data to a common grid; from 1 arcmin to 8 arcmin for the HI data by decimation, and from 7.5 arcmin to 8 arcmin by reprojection in the case of the CO data. This allows us to compare the data on a pixel-by-pixel basis.

\citet{1993ApJ...402..195W} investigated the expected CO line emission from GMCs and found that, for low UV fluxes ($1 < G_0 < 100$), the CO luminosity varies by only 10\%. In order for the CO line profiles to appear smooth and centrally peaked, a large number of clumps need to be included in the (observing) beam. Since the Taurus CO line profiles indeed show this smooth shape, we believe assuming a uniform $G_0$ is justified.

We also used the 100 $\mu$m far-infrared dust map (temperature-corrected) produced by \citet{1998ApJ...500..525S} by extracting the general Taurus area from their all-sky map. The angular resolution is 6.1 arcmin, which we smoothed and regridded to 8 arcmin to compare the results.

\section{Overview of the Taurus region}

\begin{figure}
  \resizebox{\columnwidth}{!}{\includegraphics{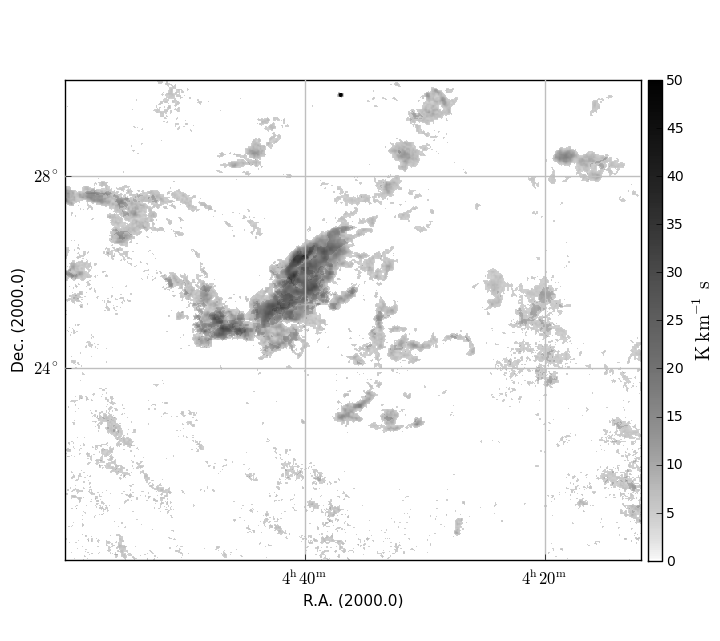}}
  \caption{\label{fig:hinsastrength} Map of the second derivative of the HI profiles indicative of the strength of HINSA features (integrated over a range of 5 to 7 km s$^{-1}$).}
\end{figure}

\begin{figure}
  \resizebox{\columnwidth}{!}{\includegraphics{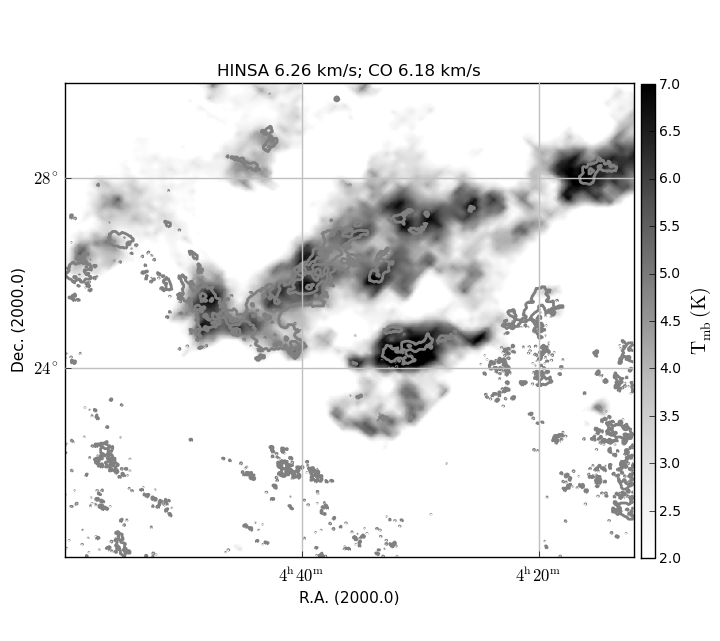}}
  \caption{\label{fig:hinsaco6} Contours of HINSA strength near 6 km s$^{-1}$\ on top of CO emission in grayscale. HINSA strength contours are 4, 12 and 20 $\rm{K~km^{-2}~s^{2}}$}
\end{figure}

In Figure \ref{fig:hico567} we show a number of CO/HI overlays that illustrate how the atomic and molecular gas in the Taurus area are intimately connected. Qualitatively, there is an anti-correlation between the CO and HI emission: the HI emission is clearly depressed in the areas where CO occurs. We aimed to locate parts of the Taurus cloud where our one-dimensional PDR model is likely to be applicable. In other words, we needed to find the parts on the map that seem to feature CO emission surrounded at the front and at the back by HI emission, along our line of sight. We then assumed that this HI emission is related to the balance of photodissociation described in the simple PDR model.

We show a fairly typical HI profile \noteA{with the corresponding CO profile (artificially magnified by a factor of 5 for clarity)} in an area dominated by CO emission in Figure \ref{fig:profile_ex}. At first glance, this profile seems to show either two overlapping HI emission peaks, or perhaps a single peak with an absorption dip at the center. Both scenarios may point to the configuration we aimed to find: atomic hydrogen surrounding carbon monoxide, or atomic hydrogen tracing dense gas by displaying a (self-)absorption feature respectively. A good fit of this particular profile can be obtained using three gaussians (shown in the figure): a broad peak, small narrow peak and one gaussian representing absorption. Poorer fits are obtained using two or three positive-only gaussians (not shown). While many pixels in the datacube of the Taurus area can be fitted through this scheme, this in itself does not necessarily imply that absorption indeed occurs. 

\citet{2003ApJ...585..823L} conducted a survey of dark clouds in the Taurus/Perseus region in search of narrow absorption features in HI and OH emission. If the HI absorption feature is narrower than the CO linewidth, they call it HI narrow self-absorption (HINSA). They found that HINSA features correlate well with molecular tracers. For the Taurus area, they found HINSA column densities of about several times $10^{18}\ \rm{cm}^{-2}$, where it should be noted that measuring the HINSA feature does not necessarily mean fully filling the `dip' in the line profile. 

Indeed, the (negative) absorption gaussian profile in our Figure \ref{fig:profile_ex} corresponds to a column density about 30 times higher than $1 \times 10^{18}~\rm{cm}^{-2}$, which would lead us to overestimate not just the absorption, but also the HI emission grossly if we were to add both the positive broad gaussian as well as the absorption gaussian to infer a total HI column density. Therefore, we decided against fitting the HI line profiles in the Taurus region as a way to identify areas where the simple PDR model can be applied.

Considering the comparatively low HINSA column densities found by \citet{2003ApJ...585..823L}, and in a more physically refined way by \citet{2008ApJ...689..276K}, relative to the overall HI column density, we finally elected to ignore the HINSA features for the purpose of calculating the HI column density related to the PDR. Instead, we opted to use the presence \note{of} HINSA only as an indication that our simple PDR model applies,
independent of the presence of any CO emission.

As mentioned in the previous Section, we assume that the main Taurus molecular cloud can be pictured as a 32-by-5 pc filament. We can use this morphological information, specifically the filament depth, to estimate a total hydrogen column density from the volume densities obtained by our method. Choosing the appropriate line of sight is important, since due to the approximative nature of the model, it is possible for it to arrive at total hydrogen column densities that are less than the HI column density. It is therefore essential to make sure the model is applied to the proper regions. In the extragalactic case this is not a problem, since the columns are sufficiently deep not to get this close to the regime where the one-dimensional PDR approximation breaks down.

\section{Atomic hydrogen column density}

\begin{figure}
  \resizebox{\columnwidth}{!}{\includegraphics{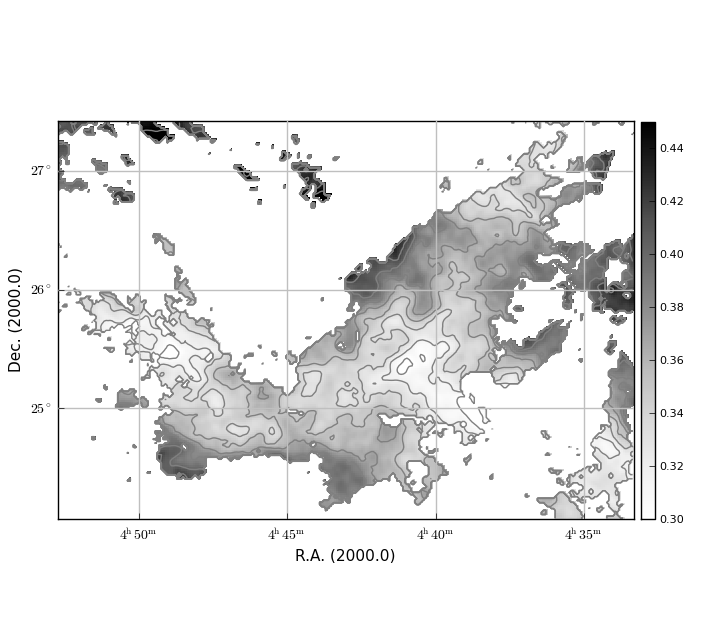}}
  \caption{\label{fig:HIfrac} Fraction of the HI column density contained in our chosen velocity range of $4 - 8.5 \rm{km s^{-1}}$ relative to the total HI column density over the range of -10 to 20 $\rm{km\ s^{-1}}$\ (both background-subtracted).}
\end{figure}

\begin{figure}
  \resizebox{\columnwidth}{!}{\includegraphics{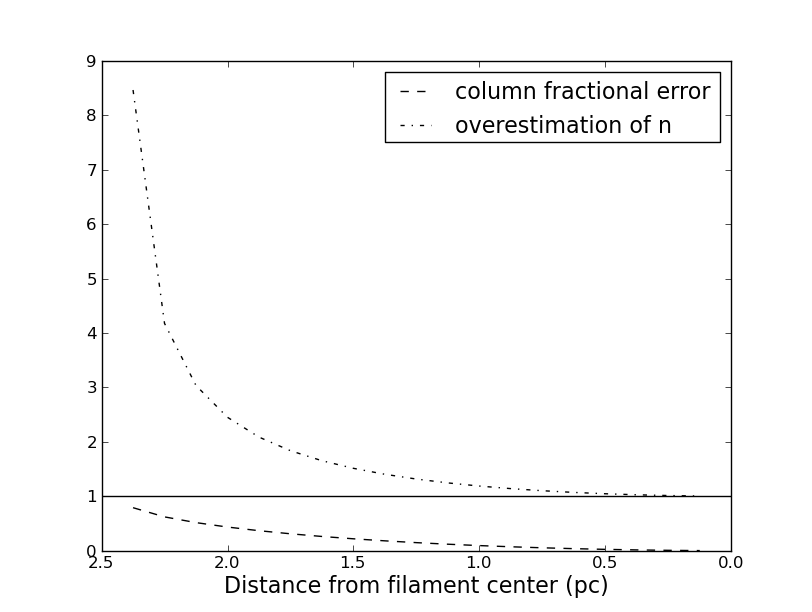}}
  \caption{\label{fig:fil_errors} Estimated errors when the HI column is not taken exactly through the center of a filament, assuming a diameter of 5 parsec. Dashed line: fractional error in the HI column, 0 at the filament center and slowly increasing going outward. Dashed-dotted line: (fractional) overestimation of the calculated total hydrogen volume density. It slowly increases up to \note{50\%} at about 2 parsec away from the center and then starts to increase rapidly.} 
\end{figure}

\begin{figure*}
  \resizebox{\columnwidth}{!}{\includegraphics{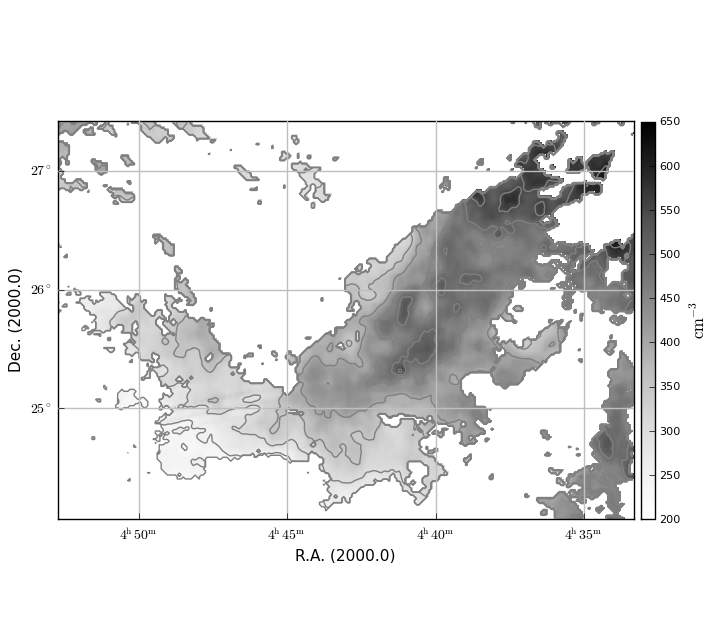}}
  \resizebox{\columnwidth}{!}{\includegraphics{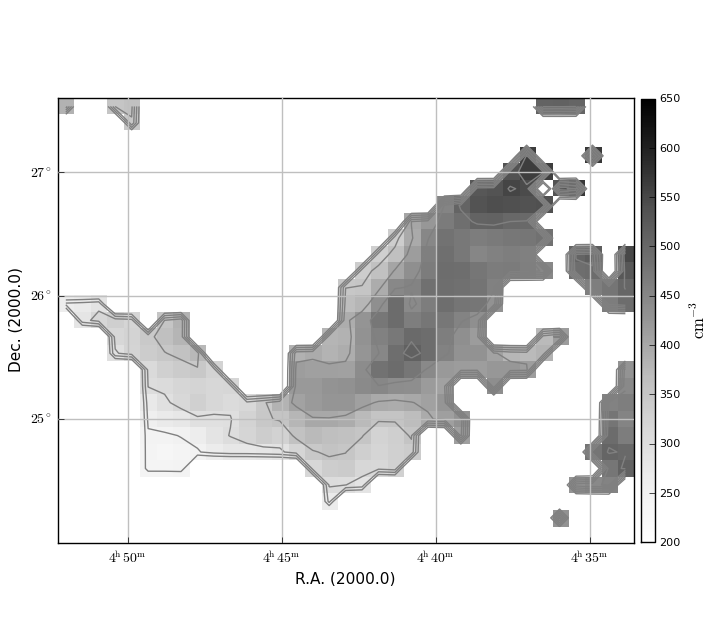}}
  \caption{\label{fig:ntotmap} Total hydrogen volume densities in the main Taurus area, full (left) and decimated (right) version (used to compare with CO emission).}
\end{figure*}


\noteA{As mentioned in the previous section,} even though we do not account for HINSA column densities in our HI calculations, we do use the presence of HINSA features to trace the pixels where we expect the simple PDR model to apply. It should be noted however that this does not mean that the model does not apply elsewhere in the cloud, so we do not necessarily trace all parts of the cloud where our model is applicable. To find HINSA, we use the method described in \citet{2008ApJ...689..276K}, where they show that HINSA is most easily seen in the second derivative of the HI line profile with respect to the velocity. The reason for this is that HINSA absorption features are typically Gaussian-like. Derivative functions of a Gaussian show a progressively stronger dependence on line width, making the narrow HINSA feature an easy-to-identify dominant feature in the second derivative as the broader overall profile is smoothed out.
We can therefore construct a map that is indicative of the strength of the HINSA feature across the Taurus region, as shown in Figure \ref{fig:hinsastrength}. We integrated the second derivative velocity channels (in units of $\rm{K~km^{-2}~s^2}$) over the range of 5 through 7 km s$^{-1}$. Then, we chose to select only those pixels with an integrated value of 4 $\rm{K\ km^{-1}\ s}$ or higher. We will refer to this integrated value as the HINSA strength.
The resulting map shows the main Taurus area as well as features in other parts of the region. Figure \ref{fig:hinsaco6} shows how the HINSA strength traces part of the Taurus CO emission.

The calculation of the atomic hydrogen column density (ignoring HINSA) requires additional assumptions. Given the very large width of the HI line ($\sim 20\ \rm{km\ s^{-1}}$; Figure \ref{fig:profile_ex}), it is natural to conclude that it originates from gas extending over much larger distances along the line of sight than the 5-pc assumed thickness for the cloud. Thus, in order to select only the atomic gas directly participating in the atomic to molecular transition into the cloud, we integrate only over the same velocity range that is spanned by the molecular gas at the position of the cloud, around 6 km s$^{-1}$, from 4.0 km s$^{-1}$ to 8.5 km s$^{-1}$ (gray area in Figure \ref{fig:profile_ex}). For our background subtraction we simply took the minimum value in each profile over the range of -10 to 20 km s$^{-1}$, and we converted the observed \note{velocity-integrated} brightness temperature $T_{b}$ to $N_{\rm{HI}}$ in the usual optically thick approximation by multiplying by $1.823 \times 10^{18}\ \rm{cm^{-2}\ \left(K\ km\ s^{-1}\right)^{-1}}$.
Considering that Taurus is located at the Galactic anti-center, we do not expect any noticeable effects of galactic rotation on the radial velocities.
A similar strategy to isolate the atomic hydrogen connected to a molecular cloud was used by e.g. \citet{2004ApJ...608..314A} and for Taurus itself by \citet{1994ApJ...433..149H}, although the latter authors used a 4 km s$^{-1}$ window centered on a 4 km s$^{-1}$ radial velocity. \note{Their HI column densities, only derived at selected positions, range from 1.6 to 2.7 $\times 10^{20}\ \rm{cm^{-2}}$. Despite the different central velocity, this is comparable to the low end of our values, which also include the higher HI column densities further away from the main Taurus region.} Figure \ref{fig:HIfrac} shows the fraction of the total HI column density (background-subtracted) we recover in our chosen velocity range as compared to integrating over the full velocity range. Typically, this fraction is between 0.3 and 0.4. 

We have also estimated the errors we make in the column and volume densities when assuming to be looking straight through a tubular filament. Our model assumes that the HI column is measured along the incident flux and, since we assume a uniform field impinging onto the surface of a filamentary (cylindrical) cloud, we should always measure the HI column straight through the center of the filament. In practice, we have a flat image and we do not know the exact filament center, so an error will be made when measuring the HI column \note{directly from} the map. Figure \ref{fig:fil_errors} shows the \noteA{error made by} taking the HI column off-axis, purely as a geometrical effect due to underestimating the path length through the filament.
Obviously, the error close to the edge of the cloud is the largest, but the error in assumed HI column density is relatively modest for most of the distance between the center and the edge of the cloud. At 2 parsecs from the center (80\% of the way) we would underestimate the HI column by about 50\% and overestimate the total hydrogen volume density by about a factor two. Considering that the HINSA features seem to occur well inside the cloud (on the plane of the sky) and not too close to the edges of the main Taurus area, we conclude that at least in the main part of the molecular cloud these errors do not dominate our results.

\section{Total hydrogen volume density maps}

\begin{figure*}
  \resizebox{\columnwidth}{!}{\includegraphics{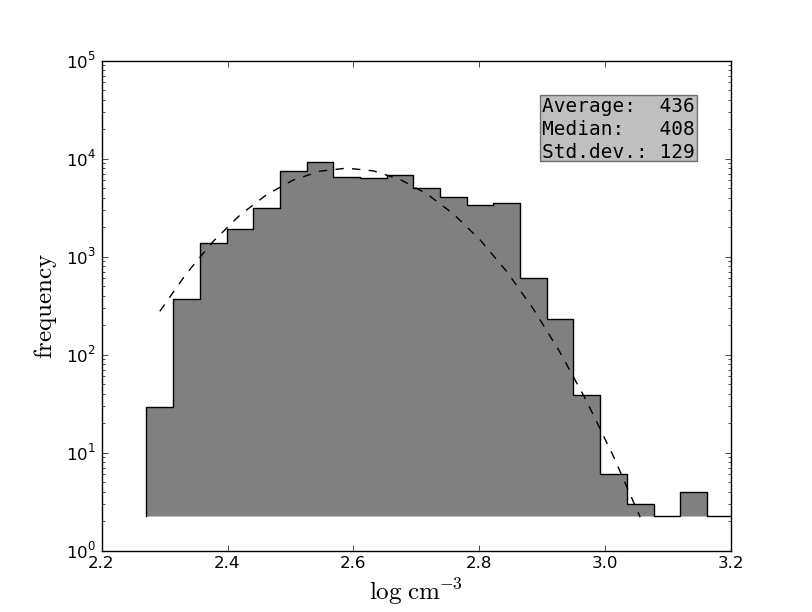}}
  \resizebox{\columnwidth}{!}{\includegraphics{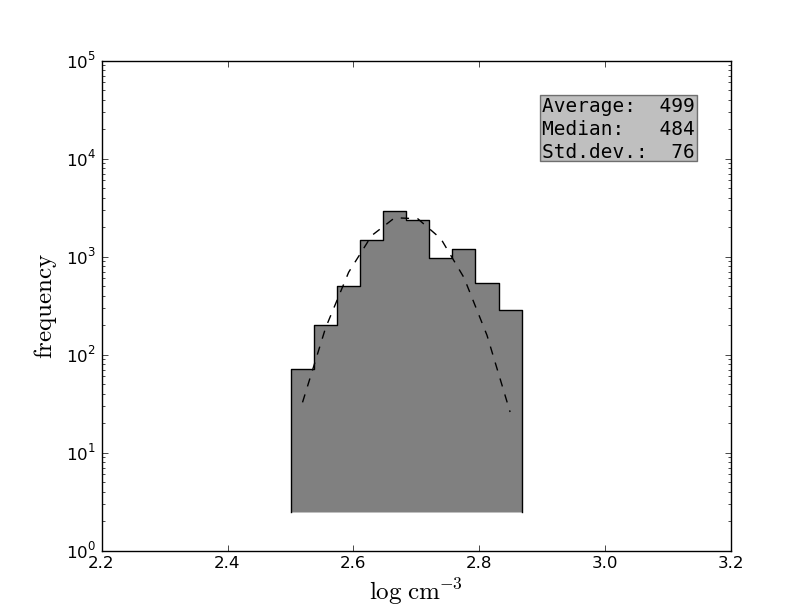}}
  \caption{\label{fig:ntotth4fits} Left panel: Distribution of logarithmic total hydrogen volume densities using a HINSA strength threshold of 4., with a simple fit of a gaussian distribution. Note that this histogram is taken from the full-resolution density map. Right panel: Same, but only for regions 2, 3, 4, and 5 defined in Section \ref{sec:CO}.}
\end{figure*}

We calculated a total hydrogen volume density map at the native HI resolution and at a reduced resolution of 8 arcmin used to match the CO data. The full resolution map and the decimated map are shown in Figure \ref{fig:ntotmap}. We stress that this calculation could be performed over a larger region. However, our model only works when we measure HI column densities close to the axis of the cloud filament -- for lines of sight approximately perpendicular to the cloud surface. We assumed that cloud filaments are traced approximately by the level of HINSA strength.

The distribution of the calculated total hydrogen volume densities is shown in the histograms in Figure \ref{fig:ntotth4fits}. In the left panel, the total hydrogen volume densities are shown for all pixels having a HINSA strength of at least 4. 
We find an average volume density of 436 $\rm{cm^{-3}}$ \noteA{and a standard deviation} $\sigma = 129~\rm{cm^{-3}}$. If we look at specific regions of the cloud with strong CO emission (next section) we obtain a more limited range of values (right panel), with densities around 500 $\rm{cm^{-3}}$. A total hydrogen volume density of about $430~\rm{cm^{-3}}$ is consistent with the $405~\rm{cm^{-3}}$ assumed by \citet{2009MNRAS.395L..81B}, although theirs is an average for the whole cloud and does not include atomic hydrogen.

The shape of the density distributions is expected to be log-normal for isothermal flows, with a power-law tail for deviations due to gravitational collapse \citep[][]{1994ApJ...423..681V,1998PhRvE..58.4501P,2008MNRAS.390..769V}. More specifically, the shape of the density distribution should evolve based on the dynamics of the cloud, and the exact value of the power-law slope depends on the projection of the cloud against the sky as well \citep[][]{2009A&A...497..399K,2011MNRAS.416.1436B,2011ApJ...727L..20K}. The left panel of Figure \ref{fig:ntotth4fits} shows a simple log-normal fit to the distribution. There is a hint of a power-law tail at the high-density end, whose existence would indicate that the high density gas of the cloud is dominated by gravity. A more sophisticated determination of the HI column density as well as a more detailed knowledge of the radiation field $G_0$ would improve the analysis of the shape of the density distribution.

Again assuming an ellipsoidal shape for this region of Taurus, we arrive at a total hydrogen mass of $\approx 1 \times 10^4 M_\odot$, \noteA{consistent with} the CO mass in \citet{1987ApJS...63..645U}.

\section{Comparison to CO emission}
\label{sec:CO}

\begin{figure*}
  \resizebox{\columnwidth}{!}{\includegraphics{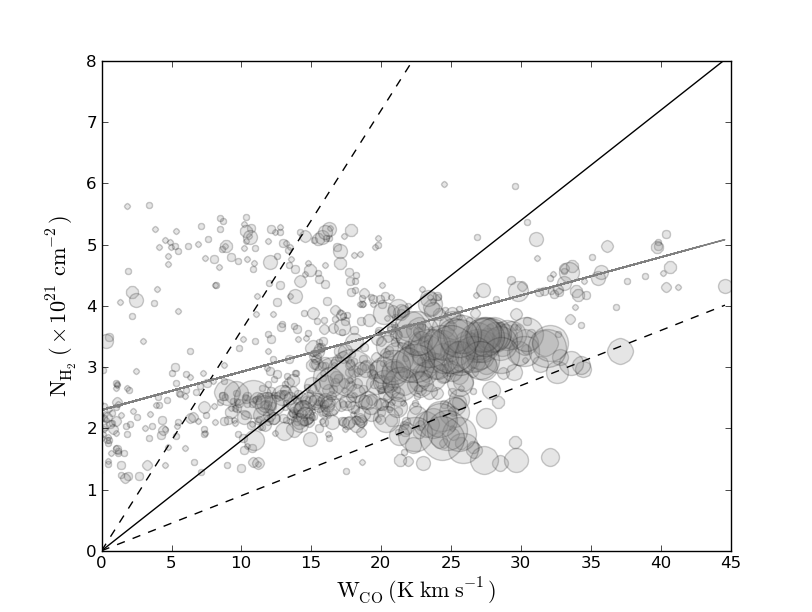}}
  \resizebox{\columnwidth}{!}{\includegraphics{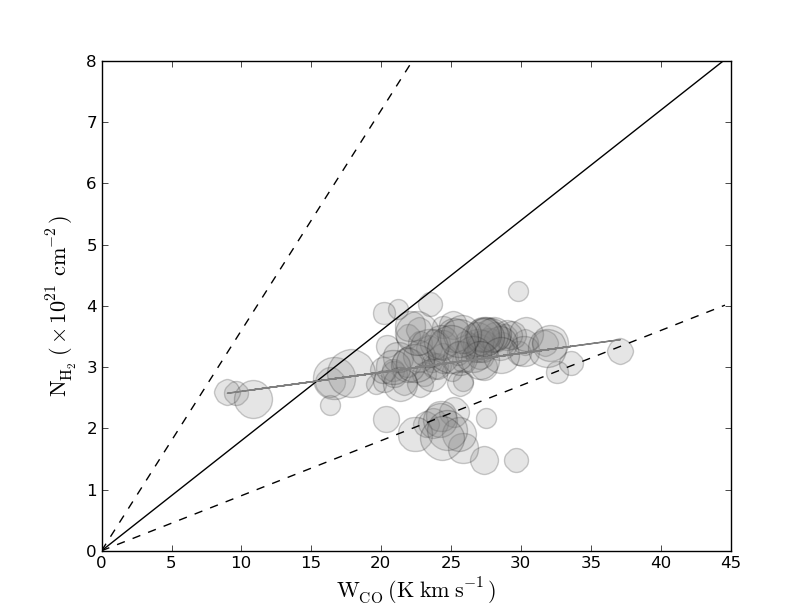}}
  \caption{\label{fig:all_th15} Comparison of predicted $N_{H_2}$ vs. CO of the full map thresholded by HINSA map values of 4 (left panel) and 15 (right panel). The black solid line is the \citet{2001ApJ...547..792D} CO X-factor. The dashed lines represent twice and half the solid line's slope. The gray solid line shows a simple least squares fit to the data points. The size of the circles is proportional to the HINSA map strength.}
\end{figure*}

Assuming a cloud depth gives us the total hydrogen column density from the total hydrogen volume density and allows us to calculate an $\rm{H_2}$ column density by subtracting the HI column density. This in turn can be compared to the CO emission from the Taurus molecular cloud. If the CO to $\rm{H_2}$ conversion factor held down to the smallest scales, then the slope of an $N_{\rm{H_2}}$ vs. $W_{\rm{CO}}$ plot would yield this CO X-factor. In practice, we would expect this relation to break down in resolved clouds because of its local fluctuations. It is further known that the X-factor also varies on a cloud-by-cloud basis \citep[e.g.][]{2007ApJ...658..446B}. For these reasons, \noteA{we caution that the} comparison between \noteA{our} $N_{\rm{H_2}}$ and $W_{\rm{CO}}$ \noteA{is for the purpose of a self-consistency check only}.


In the left and right panels of Figure \ref{fig:all_th15} we show the derived $\rm{H_2}$ column density vs. $W_{CO}$ for the pixels in the map with HINSA strength values of at least 4 and 15, respectively, using our adopted filament depth of 5 parsec. In the left panel we can distinguish roughly three features: a main cloud of points, a tail at the bottom (towards the right; more about this below) and another tail towards the upper-left part of the plot. The latter points disappear entirely for the higher HINSA strength threshold. 

We also plot a linear correlation with a slope equal to the CO factor derived by \citet{2001ApJ...547..792D} on the basis of far-infrared imaging combined with HI maps, as well as dashed lines corresponding to twice and half this value, respectively. These dashed lines correspond also approximately to the reported range of the X-factor (cf. Section \ref{sec:theory}). We see that the points fall roughly in this range, but a least-squares fit of the points themselves, shown by the solid gray line, yields an even shallower slope of $\sim 0.2$ times the \citet{2001ApJ...547..792D} X-factor.

\begin{figure}
  \resizebox{\columnwidth}{!}{\includegraphics{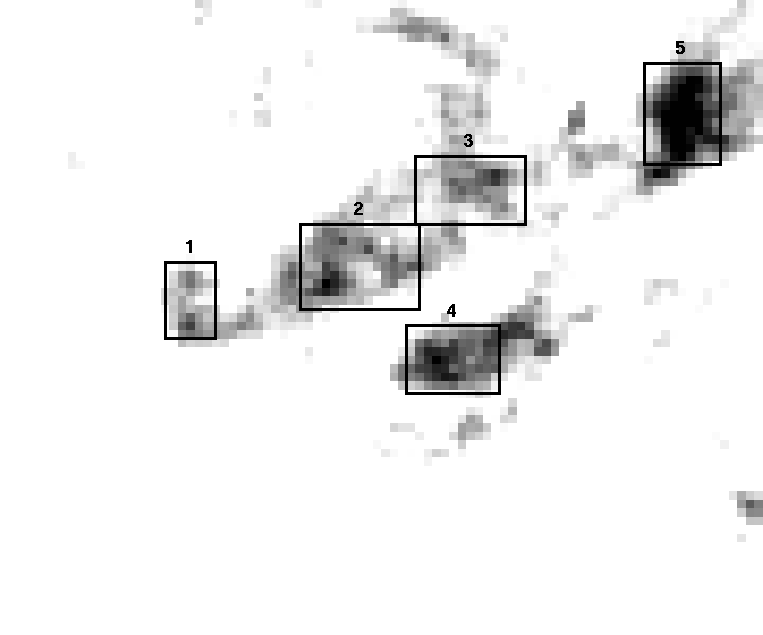}}
  \caption{\label{fig:coregions} Regions of interest (boxes) displayed on top of a CO emission map.}
\end{figure}

\begin{figure*}
  \resizebox{\columnwidth}{!}{\includegraphics{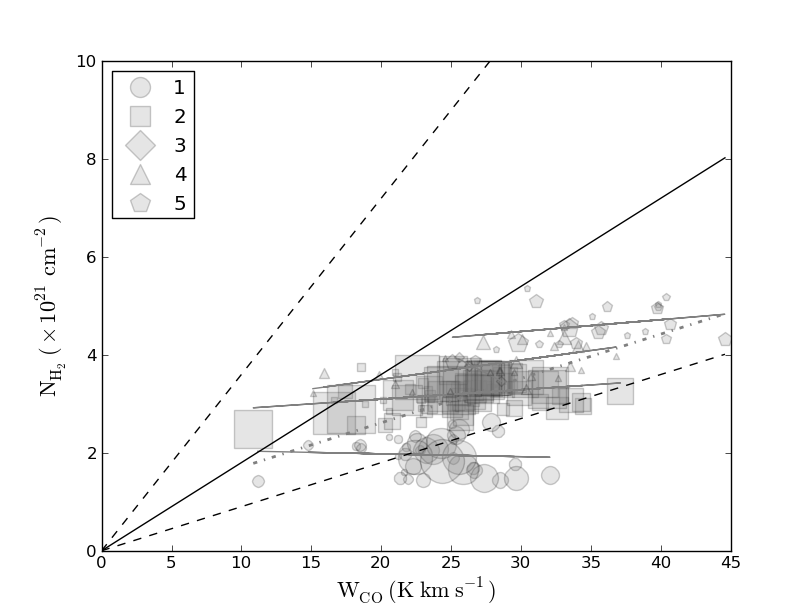}}
  \resizebox{\columnwidth}{!}{\includegraphics{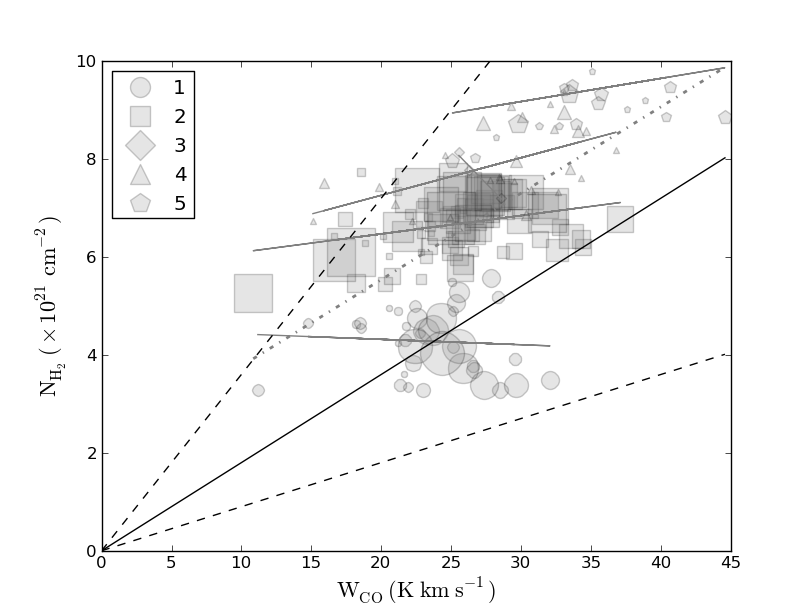}}
  \caption{\label{fig:slope_5pc} Comparison of predicted $N_{H_2}$ vs. CO of the 5 selected regions as marked on Figure \ref{fig:coregions}. A cloud depth of 5 parsec is assumed in the left panel. For each region a fitted least-square solid gray line is shown. The dash-dotted gray line is the fitted slope for all regions combined. The right panel shows the same plot, but assuming either a cloud depth of 10 parsec or $\chi=10$, since the dependency \noteA{of $N$ on $\chi$ and cloud depth} is the same. The size of the plot symbols is proportional to the HINSA map strength.}
\end{figure*}

To investigate this correlation more closely, we defined five regions of interest based on the \noteA{velocity-integrated} local CO emission, as shown in Figure \ref{fig:coregions}. These regions trace denser parts of the cloud as shown in the right panel of Figure \ref{fig:ntotth4fits}.
Figure \ref{fig:slope_5pc} (left panel) shows the pixels of these five regions \noteA{in the $\rm{N_{H_2}}$-$W_{\rm{CO}}$ plane} with different symbols. For \noteA{each one of the} regions, the slopes are shallow and lacking any real correlation, but cumulatively a slope is obtained of $0.9 \times 10^{20}~\rm{cm^{-2} (K~km~s^{-1})^{-1}}$, which is on the lower end of the range of usually quoted X-factors. 

It can be seen that the derived $N_{\rm{H_2}}$ for the points in region 1, on the eastern tip of the main Taurus area, is consistently low when compared to the observed CO emission. We speculate that our filament depth assumption is off in this area. If the filament is curved towards or away from the observer, or if the cloud resembles less of a filament in this region, then we underestimate the filament depth and as a consequence we consistently underestimate the total hydrogen \note{column} density. If the cloud is twice as thick here than assumed, this would put the points in line with the other points. Indeed, \citet{2009ApJ...698..242T} note that Taurus may indeed be as deep as it is long in certain places (25 parsecs, according to their measurement).

Region 3 does not have many points left after \noteA{imposing a} HINSA-strength \noteA{of 15}. Region 4 centers on what is thought to be a parallel filament in the Taurus cloud and assuming the same filament depth results in column densities consistent with those found in the main part. The same can be said of region 5, which is at the north-west end.

\noteA{Comparing with} Figure \ref{fig:all_th15} (left panel), we can see that \noteA{in Figure \ref{fig:slope_5pc} (left),} the points in the upper-left corner have disappeared -- they are not a part of the five selected regions in the main Taurus region. One possibility is that these points \noteA{originate from} a part of Taurus with a smaller column depth. Then our assumed column depth would be too large, leading us to overestimate total hydrogen column density and therefore also the molecular hydrogen column density. Alternatively, if the \noteA{area} to which these points belong \noteA{is part of a filamentary structure, and} has a column depth consistent with the one we assumed, the measured $N_{\rm{HI}}$ could be too far from the center of that filamentary structure, resulting in an underestimated HI column because of geometric considerations, as we showed in Figure \ref{fig:fil_errors}, and therefore an overestimated total hydrogen volume density.

We show the effect of using a different cloud depth or a different $\chi$ in the right panel of Figure \ref{fig:slope_5pc}: a depth of 10 pc is used here or a $\chi$ of 10, both yielding the same result since \noteA{$n$ is} linearly proportional to \noteA{both of them}. Recalling Equation (\ref{eqn:ntot}) and $N = n \times d_{\rm{cloud}}$, it follows immediately that $N \propto G_0 d_{\rm{cloud}}$.

It is important to stress that our results are dependent on using an accurate estimate of the incident UV flux $G_0$. Since we used one constant value, we would expect that a better model of the radiation field \noteA{would} improve the accuracy of our results.
Furthermore, we can only calculate an $\rm{H_2}$ column density if we assume a certain cloud depth.

\section{Random maps}

\begin{figure*}
  \resizebox{\columnwidth}{!}{\includegraphics{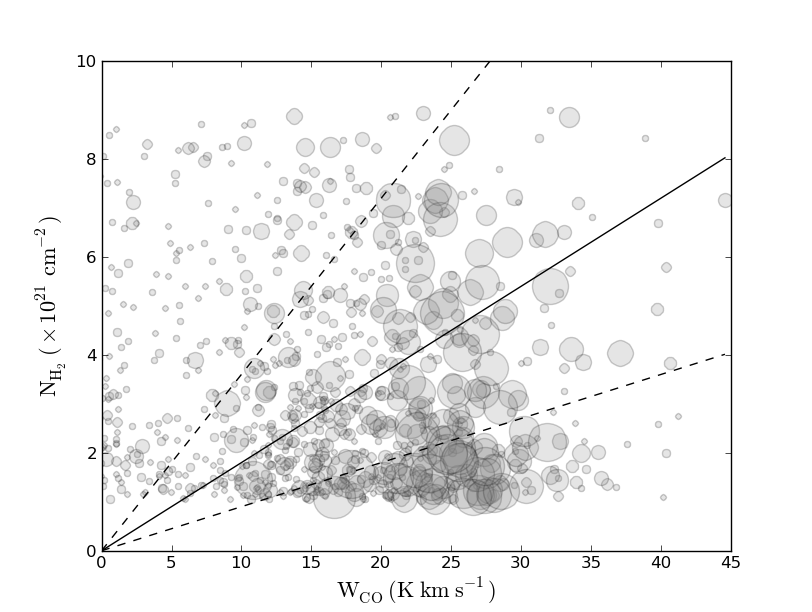}}
  \resizebox{\columnwidth}{!}{\includegraphics{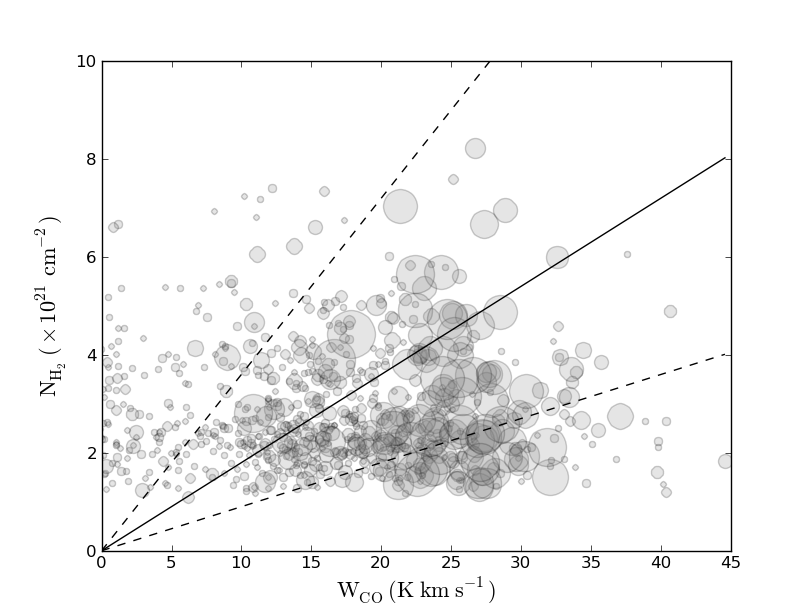}}
  \caption{\label{fig:slope_random} Left: $N_{H_2}$ vs. CO from a map of random HI column densities. Right: $N_{H_2}$ vs. CO from the GALFA-HI map with shuffled HI column densities. The circles are proportional to the HINSA strength. It can be seen that no clear correlation remains (cf. Figure \ref{fig:all_th15}).}
\end{figure*}

Because the distribution of points in Figures \ref{fig:all_th15} and \ref{fig:slope_5pc} exhibits significant amounts of scatter, it is useful to compare those results to cases in which the HI intensities in the maps are randomized, in order to see whether our results in those figures provide useful information. For this purpose, we considered the atomic hydrogen map randomized in two different ways. Then, we proceeded to calculate the total hydrogen column densities in the normal fashion, to see whether our results contain more information than these random cases.

First, we replaced all the pixels in the field by random values in the range of detected HI column densities. Based on the Taurus HI map we created a map with random column densities ranging from $2.5 \times 10^{20}$ to $9.3 \times 10^{20}~\rm{cm}^{-2}$. These values are uniformly distributed within this range.
Secondly, we took the observed HI column density map and randomly swapped pixel values. This way the global distribution of column density values is preserved, but the spatial image is destroyed.

Then, we calculated total hydrogen volume densities from both of these maps and compared the calculated $\rm{H_2}$ column densities with the real CO emission map (Figure \ref{fig:slope_random}). The point sizes are proportional to the original (non-randomized) HINSA strength at each location. It can be seen that no clear correlation is left. The left panel plot (uniformly distributed $N_{\rm{HI}}$) shows more values towards the upper part of the plot than the right panel plot (shuffled $N_{\rm{HI}}$) because the distribution of the observed HI map is not uniform and contains fewer lower-end $N_{\rm{HI}}$ values.
For comparison, see the same plot for the column densities calculated using original HI map as was shown in Figure \ref{fig:all_th15}. A much more restricted range of values is found here as well as a better correlation, and therefore we conclude that our results provide real information.

\section{Comparisons to Far-Infrared emission}

\begin{figure}
  \resizebox{\columnwidth}{!}{\includegraphics{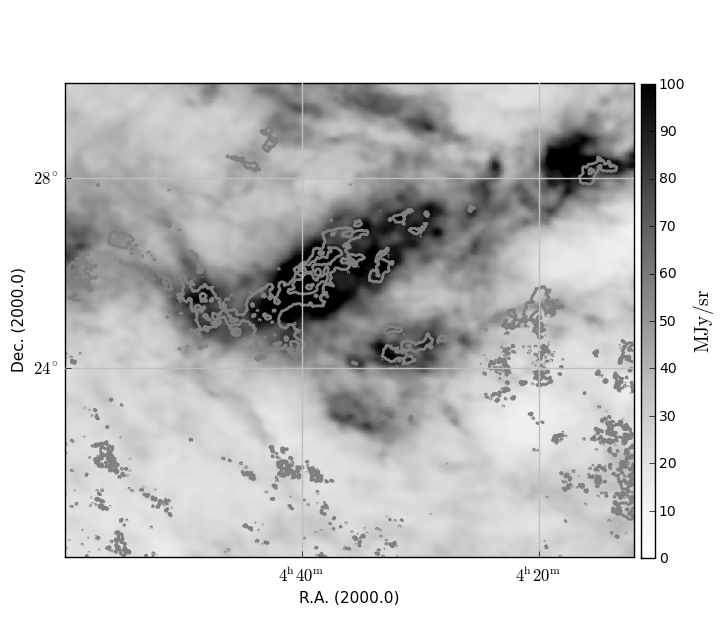}}
  \caption{\label{fig:hinsafir} Contours of HINSA strength on top of the (temperature-corrected) 100$\mu$m map. HINSA strength contours are 4, 12 and 20 $\rm{K~km^{-2}~s^{2}}$}
\end{figure}

\begin{figure*}
  \resizebox{\columnwidth}{!}{\includegraphics{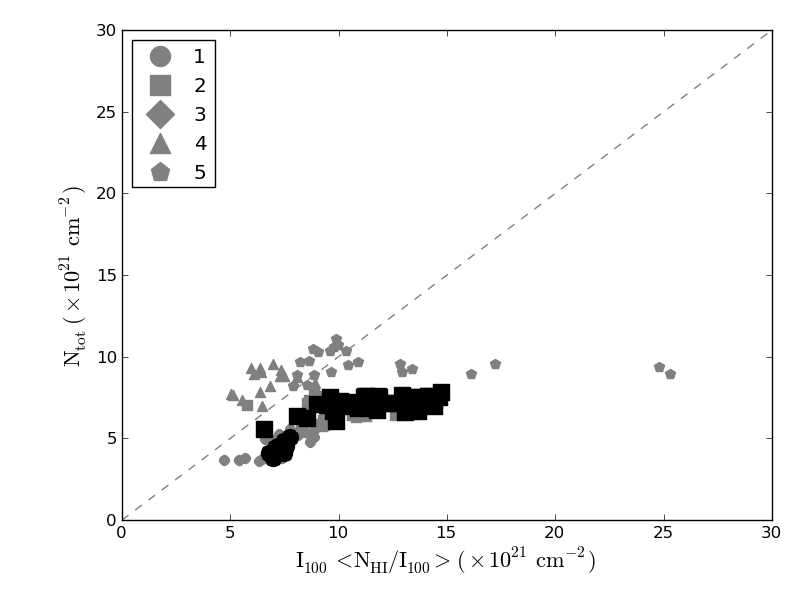}}
  \resizebox{\columnwidth}{!}{\includegraphics{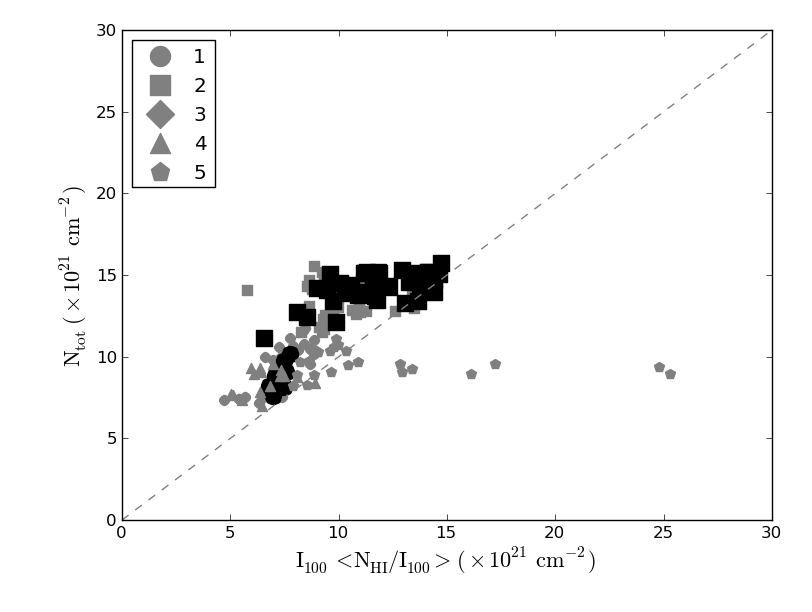}}
  \caption{\label{fig:NtotI100} Left panel: $N_{tot}$ (our method) vs. $N_{tot}$ (100$\mu$m), with a 1-to-1 line for comparison using $<N_{HI}/I_{100}> = 0.9 \times 10^{20}\ \rm{cm^{-2}/(MJy\ sr^{-1})}$. The gray dots have a HINSA strength between 4 and 15; the black dots have a HINSA strength above 15. The latter can be seen to under-predict the total gas column densities as compared to the 100$\mu$m-predicted ones. Additionally, the same regions as marked in Figure \ref{fig:coregions} are marked. Right panel: same plot but the $N_{tot}$ values of regions 1 and 2 have been artificially multiplied by 2.}
\end{figure*}

Dust emission in the far-infrared has been found to correlate well with the total hydrogen column density. \citet{2001ApJ...547..792D} used the \citet{1998ApJ...500..525S} IRAS 100 $\mu$m to calculate the CO X-factor in our own Galaxy. To this end, they calculated the ratio between the atomic hydrogen column density and the 100 $\mu$m emission in regions free of CO(1-0) emission. The resulting ratio, $0.9 \times 10^{20}\ \rm{cm^{-2}/(MJy\ sr^{-1})}$, (with an estimated uncertainty of 50\%) can then be used to calculate the total hydrogen column densities in the regions with CO(1-0) emission which in turn leads to the calculation of the CO X-factor. It should be noted that while \citet{2001ApJ...547..792D} allowed for local variations of this 100 $\mu$m conversion factor, we adopt their globally averaged ratio here.

\citet{1995A&AS..111..483A} compared the far-infrared emission of Taurus to CO emission and found that they correlate well globally. They estimate that at least 70\% of the 100 $\mu$m-traced matter is traced by CO(1-0) emission as well. In other words, a bit more gas should be traced by 100 $\mu$m emission than by CO(1-0) in the Taurus region. For reference, we plot the HINSA strength contours on top of the FIR map in Figure \ref{fig:hinsafir} (cf. Figure \ref{fig:hinsaco6}). It can easily be seen that the FIR map and the CO map are qualitatively very similar \noteA{in the densest regions. However, the infrared map shows additional diffuse structure not present in the CO map, consistent with the notion that the \note{far-infrared continuum} traces the total hydrogen gas rather than only the molecular gas.}

We compare our results to the same (temperature-corrected) far-infrared map for completeness. A plot of this comparison is shown in the left panel of Figure \ref{fig:NtotI100}. A cloud depth of 5 parsecs is assumed as before. The FIR map was smoothed and reprojected to the same resolution as the CO map, namely 8 arcmin. The gray pixels correspond to a HINSA strength of 4 to 15; the black pixels to a HINSA strength of 15 and above, and additionally the same regions as marked in Figure \ref{fig:coregions} are indicated.
There is a feature at the highest values of the 100 $\mu$m emission that is caused by a single source in the north-west part of the map (region 5) that partially \noteA{overlaps with a} HINSA detection. \noteA{This explains the outlying points on the right-hand side of Figure \ref{fig:NtotI100}.} 

It can be seen that the pixels corresponding to the higher HINSA strength are mostly confined to a single area in the plot (and regions 1 and 2). If we assume that the higher HINSA strength corresponds to a better applicability of our model, then \noteA{this suggests that the model is} underestimating the total hydrogen column density somewhat (the black pixels \noteA{indicating a HINSA strength of 15} are predominantly situated \noteA{below} the 1-to-1 correlation line). \noteA{A possible explanation of this deficiency would be that we have underestimated the column depth in region 1 and underestimated $G_0$ in region 2, both by a factor $\sim 2$ (note that $N$ is linearly proportional to both column depth and $G_0$). The right panel of Figure \ref{fig:NtotI100} shows the result of multiplying our $N_{\rm{H_2}}$ by a factor of 2 in these regions. A tighter correlation is obtained in this case.} \noteA{(See also the discussion of the radiation field in Section \ref{sec:radiation}.)}

Even though the comparison of our method to far-infrared emission shows a clear structure, if we conservatively estimate our results to have an uncertainty comparable to the infrared method, namely 50\%, \noteA{the} correlation \noteA{is significantly degraded}. \note{We expect the derived total hydrogen volume densities to have the same or better uncertainties than the ones in the nearest extragalactic case, namely M33 \citep{2011MNRAS.416....2H}}. The same range of values is obtained \note{with our method as with the far-infrared method}, meaning that both methods yield consistent results. In order to obtain better statistics a wider range of molecular clouds will need to be investigated in a similar fashion, both lower density as well as higher density clouds up to the applicability limit of $\sim 10^4\ \rm{cm}^{-3}$ total hydrogen density.

\section{Characterizing the radiation field in Taurus}
\label{sec:radiation}

\noteA{Considering the relatively low implied $\rm{H_2}$ column densities that we found in Taurus, we will discuss in this section the possibility of a stronger UV radiation field than we assumed. A stronger field would increase our inferred $\rm{H_2}$ column densities.}

\subsection{PDR model comparison}

\citet{2004ApJ...608..314A} presented \noteA{calculations \note{from}} a model of \noteA{a one-dimensional plane-parallel} PDR and how CO and HI emission relate to the incident flux $G_0$ and the total hydrogen volume density: Their Figures 1 and 2 provide diagnostic diagrams predicting a limited range of $G_0$ and $n$ given an HI column density and CO intensity. Their work was an extension of the modeling done by \citet{1999ApJ...527..795K} that solved the chemistry, radiative transfer and thermal balance of one-dimensional PDRs.
The \noteA{\citet{2004ApJ...608..314A} results predict} that the CO emission at the expected density of the Taurus region is in the regime where this emission is only very weakly dependent on the incident UV flux. \noteA{\citet{2004ApJ...608..314A} compared their calculations to the overall characteristics of G216-2.5 (Maddalena's Cloud). We will now do the same for the Taurus area.}

Taking the average CO emission over the full map yields $\langle W_{\rm{CO}} \rangle = 20.3\ \rm{K\ km\ s^{-1}}$. The average HI column density in the Taurus area in our selected velocity range is $\langle N_{\rm{HI}} \rangle = 5.7 \times 10^{20}\ \rm{cm}^{-2}$. Intersecting the values of $\langle W_{\rm{CO}} \rangle$ and $\langle N_{\rm{HI}} \rangle$ in \citet{2004ApJ...608..314A}, their Figure 2, implies $G_0 \approx 6$ and a total hydrogen volume density of several hundred $\rm{cm}^{-3}$. Computing the HI column density over a broader velocity range would raise the HI column density, implying a higher incident UV flux \noteA{if $n$ is fixed}.


Figure 20 in \citet{1991ApJ...377..192H} provides \noteA{an additional} way to infer $G_0$ from the total hydrogen volume density and the ratio $W_{\rm{CO}} / I_{\rm{100}}$. Taking the average $100\ \mu \rm{m}$ emission in the Taurus area map of 40 MJy sr$^{-1}$ we obtain $\langle W_{CO} \rangle / \langle I_{\rm{100}} \rangle = 0.5$, which for volume densities of several hundred $\rm{cm}^{-3}$ implies $G_0 \gtrsim 10$.

\noteA{The above discussion suggests that} $G_0$ \noteA{is} in the range of 1 to \noteA{perhaps} several times 10. \noteA{We consider this} quite realistic, considering the extragalactic case -- for example see \noteA{the range of $G_0$ values obtained in} M33 \citep{2011MNRAS.416....2H}. All that is needed is a cluster of a few dozen O or B stars within a few hundred parsecs distance (further than our own distance from Taurus) and a sufficiently clumpy interstellar medium to let the UV radiation penetrate. 

\subsection{Far-ultraviolet emission}

A far-ultraviolet map of the Taurus region was \note{presented} by \citet{2006ApJ...644L.181L}, although the emission is comparatively faint ($\chi \approx 0.2$). They argue that the observed FUV intensity comes mainly from foreground radiation while radiation at the back of the cloud is blocked. Far-infrared emission should not be blocked, potentially providing an alternative measure of the UV field, but they point out that the relationship between $I_{\rm{FUV}}$ and $I_{\rm{100\mu m}}$ should deviate from a simple linear relationship in optically thick regions (such as Taurus). Such a linear relationship was proposed by \citet{1995ApJ...443L..33H}, who provide their own estimate of the interstellar radiation field \citep[see also][]{2004ApJ...615..315M}. 

It is still instructive to convert the 100 $\mu$m image to an equivalent map of $\chi$ (the far-ultraviolet flux scaled by their interstellar radiation field value) using the linear relation from \citet{1995ApJ...443L..33H}. The conversion with scaling becomes:
\begin{equation}
  \chi =\left(128~I_{100} - 264\right)/{~6490}, 
  \label{eqn:I100}
\end{equation}
where $\chi$ is in equivalent dimensionless units of the usual $\chi$ since it was scaled to the interstellar radiation field value in the solar neighborhood. Scaling the 100 $\mu m$ image in this fashion yields values up to $\chi = 3$, below our assumed $\chi = 5$. However, \noteA{Figure 6 of} \citet{2004ApJ...615..315M} shows the correlation between $I_{\rm{FUV}}$ and $I_{\rm{100}}$ for a broader range of values. It can be see that the inferred values of $\chi$ can be several times higher because of significant scatter in the correlation for values similar to the ones found in the Taurus area.

\section{Conclusions}

We used a simple one-dimensional PDR model to predict total hydrogen volume densities from HI column densities in the Taurus molecular cloud in regions with strong HI narrow self-absorption (HINSA). As \citet{2004ApJ...608..314A} noted, this procedure is primarily of use in the extragalactic case. However, it is important to test the method by applying it locally in our own Galaxy in order to validate the extragalactic results. To this end, we selected atomic hydrogen column densities from a limited range of velocities around the main molecular cloud radial velocity (ignoring HINSA corrections to the column densities). 

The obtained total hydrogen volume densities appear to follow a log-normal distribution, potentially with a power-law tail at the high density end. Our method provides a promising tool to obtain the probability density function of the \noteA{volume density of the gas} in molecular clouds. We find an average total hydrogen volume density of 436 $\rm{cm}^{-3}$ ($\sigma = 129\ \rm{cm}^{-3}$) on the basis of pixels selected for HINSA features.

For the first time, we make a direct detailed comparison between our PDR-based method and CO-derived column densities \citep[cf.][]{2004ApJ...608..314A} as well as far-infrared-derived column densities. We found a \noteA{correlation} between $N_{\rm{H_2}}$ and $W_{\rm{CO}}$ \noteA{with a cumulative slope} of $0.9 \times 10^{20}~\rm{cm^{-2} (K~km~s^{-1})^{-1}}$, \noteA{implying} a relatively low value of the CO to $\rm{H_2}$ conversion factor, at least when the full ensemble of data points is considered (individual parts of the Taurus cloud \noteA{essentially} show no correlation. However, \noteA{assuming that the far-infrared emission is a good tracer of the total gas, our method yields lower total hydrogen column densities} as compared to the \noteA{column densities implied by the} far-infrared emission, at least for higher values of the HINSA strength. There are three \noteA{ways} in which our total hydrogen column densities \noteA{can be made larger: using} a larger cloud depth, a higher incident UV flux $G_0$, or a lower HI column density (obtained by narrowing the velocity range of atomic hydrogen associated to Taurus).

We adopted $\chi=5$ ($G_0 = 4.25$) from \citet{2010ApJ...715.1370G}, although in line with their findings, this seems to be the lowest value for the UV radiation field that fits the observations. \noteA{Twice} this value might yield better results.  \noteA{We conclude, based on our results and the available literature, that a higher $G_0$ of at least 10 in the general Taurus area would be expected}.

In the case of the central region of Taurus, it appears that all the gas mass is accounted for \noteA{(since our results are consistent with both CO and FIR observations)}, unless the true $G_0$ pervading the molecular cloud is, unexpectedly, \noteA{at least several times higher than previously assumed}. \noteA{We might expect to find hidden `dark' gas based on the work by \citet{2010ApJ...716.1191W}, or the observations of galactic supershells by \citet{2011ApJ...728..127D}.} 
The parts of the Taurus area where we applied the \noteA{plane-parallel} PDR model were not CO-selected, \noteA{but rather by the presence of HINSA; however,} all selected areas do have detectable CO emission. One could assume that molecular gas not traced by CO is present in the atomic hydrogen layer surrounding the CO cloud, but the results of our method show no evidence of this in the form of predicted total hydrogen column densities consistently higher than those inferred by CO emission or infrared emission. Molecular gas not traced by CO could still be present in the areas directly surrounding Taurus, further away from the CO cloud on the plane of the sky. However, the lack of better information about the detailed UV radiation field in the Taurus region prevents a thorough search at this time. It would be of interest to apply the same model to similar clouds in our Galaxy with hints of `dark' gas that is not traced by CO or where higher X-factors have been inferred.

Assuming a consistent relation between our inferred molecular gas versus the CO-implied gas, we speculate that the eastern tip of the Taurus cloud has twice the thickness of the main area, independent of what thickness is adopted or will be found for the main area. 

The results of this work could be improved through a combination of cloud depth information to aid the direct comparison to CO \citep[anticipating e.g. the Gould Belt Distance Survey,][]{2011RMxAC..40..205L} and a more sophisticated way to calculate local (related) HI column densities. \noteA{More accurate HI column density measurements would require, for example, a way to distinguish foreground and background HI emission from emission from the molecular cloud, if we could find out more about the local velocity and temperature structure.}

Going forward, we expect to be able to use the simple PDR model \noteA{in future work} as a tool for investigating the atomic-molecular transition in molecular clouds.

\section*{Acknowledgments}
JSH acknowledges constructive comments from Ron Allen about the application of the simple PDR model and from Laurent Loinard about the morphology of the Taurus molecular cloud that helped improve the results. JSH is supported by CONACYT grant 102488 to EVS. We thank the anonymous referee for suggestions that improved the clarity of this work, particularly in terms of clarifying the differences between the Galactic and the extragalactic case.

\label{lastpage}
\bibliographystyle{mn2elong} 
\bibliography{references}

\end{document}